\documentclass{article}
\usepackage{graphicx} % Required for inserting images
\usepackage{amsmath,amsfonts,amssymb, amsthm}
\usepackage{setspace}
\usepackage{tocloft}
\usepackage{hyperref}
\usepackage{float}
\usepackage{authblk}
\usepackage{algorithm}
\usepackage{algpseudocode}
\usepackage{float}
\usepackage{pgfplots}
\pgfplotsset{compat=1.18}
\usepackage{pgfplotstable}

\newtheorem{definition}{Definition}

\providecommand{\keywords}[1]
{
  \small	
  \textbf{Keywords: } #1
}

\title{More Efficient Stealth Address Protocol}
\author[1]{Marija Mikić}
\author[2]{Mihajlo Srbakoski}
\author[3]{Strahinja Praška}
\affil[1,2 ]{Faculty of Mathematics, University of Belgrade}
\affil[3]{Faculty of Technical Sciences, University of Novi Sad}
\date{April 2025}

\begin{document}

\maketitle

\begin{abstract}

The integration of privacy-preserving transactions into public block\-chains such as Ethereum remains a major challenge. The Stealth Address Protocol (SAP) provides recipient anonymity by generating unlinkable stealth addresses. Existing SAPs, such as the Dual-Key Stealth Address Protocol and the Curvy Protocol, have shown significant improvements in efficiency, but remain vulnerable to quantum attacks. Post-quantum SAPs based on lattice-based cryptography, such as the Module-LWE SAP, on the other hand, offer quantum resistance while achieving better performance.

In this paper, we present a novel hybrid SAP that combines the Curvy protocol with the computational advantages of the Module-LWE technique while remaining Ethereum-friendly. In contrast to full post-quantum solutions, our approach does not provide quantum security, but achieves a significant speedup in scanning the ephemeral public key registry, about three times faster than the Curvy protocol. We present a detailed cryptographic construction of our protocol and compare its performance with existing solutions. Our results prove that this hybrid approach is the most efficient Ethereum-compatible SAP to date.

\end{abstract}


\section{Introduction}

There is a strong demand for the integration of private transactions into public block\-chains such as Ethereum \cite{eth}. Various techniques can be used to protect transaction details, including the amount, sender and recipient address. The best-known methods include ring signatures, zero-knowledge proofs and stealth addresses. By using stealth addresses, we can ensure the privacy of the recipient of the transaction. For certain applications, this is a sufficient feature of the three mentioned.

Different variants of the stealth address protocol can be used. Some, such as DKSAP \cite{toni} (Dual-Key Stealth Address Protocol), are based on a cryptographic mechanism similar to the Diffie-Hellman method \cite{dh}. In this approach, the private and public keys for the stealth address contain a shared secret that can be derived from both the sender and the recipient. The shared secret is computed by performing scalar multiplication of an elliptic curve point. However, the dual-key stealth address protocol can be generated using a different cryptographic technique, namely elliptic curve pairing. Several studies have addressed the development of a Stealth Address Protocol (SAP) based on pairing, including works \cite{kinezi1, kinezi2}. However, in \cite{kinezi1}, a vulnerability is uncovered that allows the sender and an entity with access to the viewing key to jointly determine the private key of the stealth address. In contrast, \cite{kinezi2} proposes that the private key of the stealth address can be computed by the sender alone, without the need for collaboration. This is the motivation for our paper \cite{MM}. It overcomes all the shortcomings mentioned above and creates new protocols based on elliptic curve pairing. Protocol 3 from the paper \cite{MM} (later called Curvy protocol \cite{curvy}) is Ethereum-friendly and provides five times higher efficiency than the fastest DKSAP work based on elliptic curve multiplication \cite{toni}.

All previously discussed DKSAPs rely on the elliptic curve discrete logarithm problem, which can be efficiently solved by a sufficiently powerful quantum computer using Shor’s algorithm. In our paper \cite{PQMM}, we introduced three new post-quantum stealth address protocols (SAPs) based on lattice-based cryptography:\ LWE SAP, Ring-LWE SAP and Module-LWE SAP. These protocols utilize the Learning With Errors (LWE) problem to provide quantum-resistant privacy. Among them, Module-LWE SAP, which is built upon the Kyber key encapsulation mechanism, demonstrates the best performance, surpassing the Curvy protocol by approximately 66.8\% in the scan time of the ephemeral public key registry.

In this paper, we want to combine the Curvy protocol and the Module-LWE SAP to create a DKSAP that is Ethereum-friendly but even more faster than the Curvy protocol. We want to utilise the efficiency that the Module-LWE technique provides, but the resulting protocol does not provide post-quantum (PQ) security. This way we get the best performing SAP protocol that is Ethereum friendly. The results obtained show that the new protocol is about 3 times faster than the Curvy protocol (in terms of the time needed to scan the ephemeral public key register).

The paper consists of six sections. The first section is the introduction. The second section discusses protocols that already exist and that were the motivation for our work. Section 3 introduces us to lattice-based cryptography. We define the concept of lattice, ideal lattice, Shortest Vector Problem (SVP) and Short Integer Solution (SIS). Then we focus on the Module-LWE problems. We define both the Search-MLWE problem and the Decision-MLWE problem. Section 4 is devoted to our proposed protocol, which uses MLWE and elliptic curve multiplication and has two private keys. This section consists of three subsections. The first one is the background, the second one is a brief description of the Kyber encapsulation mechanism and the last one is dedicated to our protocol. In section 5 we have two subsections. The first one is a comparison of our protocol with the MLWE SAP. The second is a comparison of our protocol with the Curvy protocol. The last section is the conclusion.

\section{Related work}

The development of stealth address protocols has been an area of research in blockchain privacy since their introduction in 2011. Early developments, such as the Dual-Key Stealth Address Protocol (DKSAP), laid the foundation for later advances in efficiency and security. The introduction of viewing keys in DKSAP enabled selective disclosure, but the protocol remained computationally expensive, especially on low-performance devices. Various improvements were proposed, including optimizations using elliptic curve pairing to reduce the scan time of the ephemeral public key registry.

Several works have addressed the construction of pairing-based stealth address protocols. Paper \cite{MM} introduced ECPSKSAP (Elliptic Curve Pairing Single-Key Stealth Address Protocol) and three variants of ECPDKSAP (the Protocol 3 is later called Curvy protocol \cite{curvy}). Implementation results \cite{MM} showed that the Curvy protocol achieved up to a 80\% improvement in scanning efficiency compared to DKSAP \cite{toni}.

In parallel, privacy-preserving techniques, such as the Zerocash protocol \cite{textbook7}, have introduced zero-knowledge techniques to hide the identities of sender and recipient. While the Zerocash-based protocols Railgun \cite{textbook10} and Labyrinth \cite{textbook11} offer strong anonymity guarantees through shielded transactions, DKSAP-based systems such as Umbra Cash \cite{textbook8}, Fluidkey \cite{textbook9} and Curvy \cite{curvy} offer weaker security guarantees that only ensure the privacy of the recipient. However, they are significantly faster and more cost-efficient.

Given the growing threat of quantum computing, interest in developing post-quantum stealth address protocols has increased. Some early efforts at PQ SAPs include the use of fully homomorphic encryption \cite{fhedksap}, but this approach relies on elliptic curve cryptography and does not provide true quantum resistance.

Lattice-based cryptographic primitives, especially those based on LWE, Ring-LWE and Module-LWE, have proven to be strong candidates for the construction of post-quantum stealth address protocols. FrodoKEM \cite{frodo-kem-paper}, NewHope \cite{newhope-paper} and Kyber \cite{kyber} have been proposed as secure key encapsulation mechanisms and provide a foundation for PQ SAP designs \cite{PQMM}. Our work builds on these findings by integrating Module-LWE techniques into the Curvy protocol to further optimize stealth address protocol.

\section{Lattice-based Cryptography}

In this section, we provide an overview of some of the foundational problems in the post-quantum cryptography, focusing on the LWE problems and its structured variants, Ring-LWE and Module-LWE. These problems not only underpin the security of many cryptographic constructions but also offer efficient implementations due to their algebraic structures. The foundational importance of lattice problems in cryptography was first highlighted by Ajtai, who demonstrated the hardness of generating lattice problems with cryptographic relevance, in his paper \cite{ajtai}. His results laid the groundwork for developing lattice-based cryptographic schemes, including those based on the LWE problems.

We begin by introducing lattices and ideal lattices, SVP and SIS, which form the mathematical basis for these problems, and then proceed to the LWE, Ring-LWE and Module-LWE problems.

\subsection{Notation and basic definitions}

In our protocols, we use a ring of integers $\mathbb{Z}$ and the polynomial ring $R = \mathbb{Z}[x]/(x^n + 1)$, where $n$ is a power of 2. For an integer $q$, we use $\mathbb{Z}_q = \mathbb{Z}/q\mathbb{Z}$ to denote the ring of integers modulo $q$, and $R_q$ to denote $R/qR=\mathbb{Z}_q[x]/({x^n+1})$. We denote by regular letters elements in $R, R_q, 
\mathbb{Z}, \mathbb{Z}_q$, with lowercase bold letters vectors from $R_q^k$ (where $k>1$) and with uppercase bold letters matrices from $R_q^{k \times k}$ (where $k>1$).

Following definitions (see \cite{lattice5}) of a lattice and its basis, provide a formal framework for understanding lattice-based problems.

\begin{definition}
Let $\mathbb{R}^m$ be the $m$-dimensional Euclidean space. A lattice in $\mathbb{R}^m$ is the set 
$$
\mathcal{L}(\mathbf{b}_1, \dots, \mathbf{b}_n) = 
\left\{
\sum_{i=1}^n x_i \mathbf{b}_i : x_i \in \mathbb{Z}
\right\}
$$
of all integral combinations of $n$ linearly independent vectors $\mathbf{b}_1, \dots, \mathbf{b}_n$ in $\mathbb{R}^m$ (where $m \geq n$). The integers $n$ and $m$ are called the rank and dimension of the lattice, respectively. The sequence of vectors $\mathbf{b}_1, \dots, \mathbf{b}_n$ is called a lattice basis.
\end{definition}

A lattice is also characterized as an additive subgroup of $\mathbb{R}^m$ that is discrete, meaning it has no accumulation points other than infinity. This implies that the lattice consists of isolated points and forms a regular grid-like structure in the vector space. 

To define an ideal lattice, we first introduce the concept of an ideal.

\begin{definition} An ideal $\mathcal{I}$ of the ring $R$ is an additive subgroup $\mathcal{I} \subseteq R$ that is closed under multiplication by elements of $R$, meaning $v \cdot r \in \mathcal{I},$ $\forall v \in \mathcal{I}$ and $\forall r \in R$.
\end{definition}

An ideal lattice is a lattice derived from an ideal $\mathcal{I}$ in a commutative ring $R$ under a specific embedding, such as the coefficient embedding.  This multiplicative property introduces unique structural characteristics to ideal lattices.

Lattice-based cryptography relies on several hard computational problems. Among these, some are particularly relevant for their direct application in cryptographic schemes, and we will now define three of them (the definitions are taken from \cite{lattice3}, where one can also find definitions for the Decisional Approximate SVP (\(\text{GapSVP}_\gamma\)), Approximate Shortest Independent Vectors Problem (\(\text{SIVP}_\gamma\)), and Bounded Distance Decoding Problem (\(\text{BDD}_\gamma\))). 

The first is the Shortest Vector Problem (SVP), which is central to understanding lattice structures and their cryptographic utility.

\begin{definition}\textbf{(SVP)}
Given an arbitrary basis \( \mathbf{B} \) of some lattice \( \mathcal{L} = \mathcal{L}(\mathbf{B}) \), find a shortest nonzero lattice vector, i.e, a vector \( \mathbf{v} \in \mathcal{L} \) for which $$ \|\mathbf{v}\| = \min_{\mathbf{u} \in \mathcal{L} \setminus \{\mathbf{0}\}} \|\mathbf{u}\| .$$
\end{definition}

In cryptography, approximation variants of lattice problems are often more practical due to their reduced computational complexity. These problems are parameterized by an approximation factor \( \gamma \geq 1 \), which is typically a function of the lattice dimension $n$. For instance, the approximation version of SVP is known as \( \text{SVP}_\gamma \).

\begin{definition}\textbf{ (\( \text{SVP}_\gamma \))}
Given a basis \( \mathbf{B} \) of an \( n \)-dimensional lattice \( \mathcal{L} = \mathcal{L}(\mathbf{B}) \), find a nonzero vector \( \mathbf{v} \in \mathcal{L} \) for which $$\|\mathbf{v}\| \leq \gamma(n) \cdot \min_{\mathbf{u} \in \mathcal{L} \setminus \{\mathbf{0}\}} \|\mathbf{u}\|. $$ 
\end{definition}

Another crucial problem in lattice cryptography is the Short Integer Solution (SIS) problem. Informally, the SIS problem asks us to find a "short" nonzero integer combination of given lattice vectors that sums to zero. 

\begin{definition}\textbf{ (\( \text{SIS}_{n,q,\beta,m} \))}
Given \( m \) uniformly random vectors \( \mathbf{a}_i \in \mathbb{Z}_q^n \), forming the columns of a matrix \( \mathbf{A} \in \mathbb{Z}_q^{n \times m} \), find a nonzero integer vector \( \mathbf{z} \in \mathbb{Z}^m \) of norm \( \|\mathbf{z}\| \leq \beta \) such that 
\[
 \mathbf{A} \mathbf{z} = \sum_i \mathbf{a}_i \cdot z_i = 0 \in \mathbb{Z}_q^n.
\] 
\end{definition}

\subsection{LWE and Ring-LWE problems}  

The Learning With Errors (LWE) problem, introduced by Regev \cite{regev}, is a fundamental post-quantum hardness assumption based on the difficulty of distinguishing noisy linear equations from random ones. Its structured variant, Ring-LWE (RLWE), improves efficiency by replacing vectors and matrices with polynomial ring elements. Both problems underpin secure encryption and digital signatures.

Peikert \cite{lattice3} highlights the duality between LWE and the Short Integer Solution (SIS) problem, reinforcing LWE’s security against classical and quantum attacks. Regev \cite{regev} further proves that solving LWE efficiently would also solve the Shortest Vector Problem (SVP) in structured lattices, a problem believed to be intractable.

The Ring-LWE problem, introduced by Lyubashevsky, Peikert, and Regev \cite{regev2}, extends these guarantees to ideal lattices. Their Main Theorem 1 states that if SVP is hard for quantum algorithms, then Ring-LWE samples appear pseudorandom to any polynomial-time (possibly quantum) adversary, ensuring both security and efficiency in cryptographic applications.

A detailed formalization of LWE and Ring-LWE problems, along with their applications in post-quantum stealth address protocols, can be found in \cite{PQMM}.

\subsection{Module-LWE problems}

In the paper \cite{fhe-gentry} is introduced the General Learning With Errors (GLWE) problem, for the purpose of constructing a fully homomorphic encryption scheme without bootstrapping, which was the first step towards considering the Module-LWE problem.

Module-LWE is explained in detail in the paper \cite{module}. This work extended the foundational principles of LWE and Ring-LWE to module structures, providing a versatile framework that enables the development of cryptographic schemes combining both efficiency and adaptability. As in the case of LWE and Ring-LWE problems, we define distributions, as well as the associated search and decision problems. These definitions are based on the work in \cite{module}.

\begin{definition}
\textbf{(Module-LWE Distribution)} For an $\textbf{s} \in R_q^k$ called secret, the Module-LWE distribution $A_{\textbf{s}, \chi}$ over $R_q^k \times R_q$ is sampled by choosing $\textbf{a} \in R_q^k$ uniformly at random, choosing $e $ from $ \chi,$ and outputting $\textbf{a}$ and $b = ( \langle \textbf{s}, \textbf{a} \rangle + e) \mod q.$
\end{definition}

\begin{definition} \hspace{-2.5mm}
\textbf{(Search-MLWE}$_{q,\chi,m,k}$\textbf{)} \hspace{-1.5mm} Given $m$ independent samples \mbox{$(\textbf{a}_i, b_i) \! \in \! R_q^k \! \times \! R_q$} from $A_{s, \chi}$ for a uniformly random $\textbf{s} \in R_q^k$ (fixed for all samples), find $\textbf{s}$.
\end{definition}

\begin{definition} \hspace{-2.8mm}
\textbf{(Decision-MLWE}$_{q,\chi,m,k}$\textbf{)} \hspace{-1.7mm} For $m$ independent samples \mbox{$(\textbf{a}_i, b_i) \! \in \! R_q^k \! \times \! R_q$}, where every sample is distributed according to either:  (1) $A_{\textbf{s}, \chi}$ for a uniformly random $\textbf{s} \in R_q^k$ (fixed for all samples), or (2) the uniform distribution, distinguish which is the case (with non-negligible advantage).
\end{definition}

Module-LWE generalizes Ring-LWE by working with module over a polynomial ring instead of just the rings themselves, allowing for more flexible dimensions and balancing security and efficiency. It inherits the hardness assumptions of both LWE and Ring-LWE, under the appropriate parameter settings. Its flexibility and efficiency have made it a foundation for modern PQ cryptographic schemes, such as Kyber \cite{kyber} and Dilithium \cite{dilithium}, which are finalists in the NIST PQ standardization process.

\section{Module-LWE SAP for Blockchain Applications}

In this section, we provide an overview of Kyber and its integration into the Efficient Curvy protocol. The focus of this section is the Efficient Curvy protocol, as it represents the most optimized version of our SAP. Before delving into the protocol itself, we first introduce the necessary background and define the notation used throughout this section.

\subsection{Background}

We begin this section with the modular reduction and define infinity norms in the corresponding spaces, which we will use in the rest of the paper.

\begin{definition}\textbf{\mbox{(Modular reduction)}}
    For any positive integer $q$ and $r \in \mathbb{Z}$, we define $r' := r \bmod{q}$, $ r'\in \mathbb{Z}_q$, to be a unique element such that $0 \leq r' < q$.
\end{definition}

\begin{definition}\textbf{(Symmetric modular reduction)}
For an even positive integer $q$ and $r \in \mathbb{Z}$, we define 
$$\displaystyle
r' := r \bmod^+{q} =
\begin{cases} 
r & \text{if } r \leq \frac{q}{2}, \\
r-q   & \text{if } r > \frac{q}{2} \\
\end{cases} 
.$$
\end{definition}

For odd $q$, we would have a case that $\frac{q-1}{2} \leq r' < \frac{q-1}{2}$, everything else stays the same. 
Note that newly defined $r'$ belongs to $\mathbb{Z}_q$.

For an element $r \in \mathbb{Z}_q$, we define $\lVert r \rVert_\infty := \lvert r \bmod^+{q} \rvert$. For $r \in R_q$ we define it as $\lVert r \rVert_\infty := \max\limits_{i} \lVert r_i \rVert_\infty$, where $r_i$ are coefficients of polynomial $r$ in $R_q$, i.e.\ $r_i \in \mathbb{Z}_q$. For $\textbf{r} = (r_1, r_2, ..., r_k)  \in R_q^k$, we define it as $\lVert \mathbf{r} \rVert_\infty := \max\limits_{j} \lVert r_j \rVert_\infty$, where $r_j \in R_q$.

In the following, we introduce some definitions that are necessary due to the concepts we use in the rest of the paper and their notation.

We use $\lceil \cdot \rceil $ and   $\lceil \cdot \rfloor$
to denote ceiling and rounding to the
nearest integer functions (rounding down in the case of a tie).

\begin{definition}\textbf{(Centered binomial distribution)}  
Let $\eta$ be a positive integer. The centered binomial distribution $B_\eta$ is defined as the probability distribution over  
\[ 
x = \sum_{i=1}^\eta (a_i - b_i),  
\]  
where $a_i, b_i \sim \text{Ber}(0.5)$, for $i \in \{1, 2, \dots, \eta\}$, are independent random variables sampled from the Bernoulli distribution with parameter $p = 0.5$.  
\end{definition}  

Note that the output $x$ belongs to the set $\{-\eta, -\eta+1, \dots, \eta-1, \eta\}$.

In our protocols vectors of polynomials, as well as matrices of polynomials, from $R_q$ and $B_\eta$, where $B_\eta$ is a centered binomial distribution with parameter $\eta$. In the Kyber \cite{kyber} key encapsulation mechanism, a centered binomial distribution $B_\eta$ is used for $\eta=2$ and $\eta=3$.

When we write that a polynomial is from $B_\eta$, this means that each coefficient is sampled from $B_\eta$.  A vector of polynomials from $R^k$ can be sampled from $B^k_\eta$.

Extendable Output Function ($\text{XOF}$) is a function on bit strings where the output can be extended to any desired length. If we want $\text{XOF}$ to take $x$ as input and then produce a value $y$ that is distributed uniformly over a set $T$, we write as $y \sim T : = \text{XOF}(x)$. This function is deterministic, i.e.\ it always generates the same $y$ for a given $x$. If we write $\textbf{A} \sim R_q^{k \times k} := \text{XOF}(x)$, we mean that the output of XOF is $\textbf{A} \in R_q^{k \times k}$.

\subsection{Kyber}  

Kyber is a secure post-quantum Key Encapsulation Mechanism (KEM) based on the Module-LWE problem. It is one of the finalists in the NIST competition for post-quantum cryptography.

There is a Kyber512, a Kyber768 and a Kyber1024 version, each with different security levels. Kyber512 has a security level roughly equivalent to AES128, Kyber768 is roughly equivalent to AES192 and Kyber1024 is roughly equivalent to AES256.

We use Kyber to calculate the shared secret $S$ (both sender and recipient calculate it) in the protocol in the subsection Efficient Curvy Protocol. Thanks to the performance of Kyber compared to the operation on the elliptic curve, we optimize the search of the Ephemeral public key registry in this way.

\subsubsection{Compression and Decompression}  

To optimize parameter sizes, Kyber uses compression and decompression functions, which discard low-order bits of public keys and ciphertexts. Given a modulus \( q \) and bit-length parameter \( d \), these functions map elements between \( \mathbb{Z}_q \) and a smaller range \( \mathbb{Z}_{2^d} \), reducing storage and transmission costs.  

Decompression approximates the original value, introducing a bounded error. Specifically, if \( x' \) is obtained by applying decompression to a compressed value of \( x \), then $x'$ is close to $x$, i.e.\ 
\[
\lvert (x' - x) \bmod^+ q \rvert \leq \left\lceil \dfrac{q}{2^{d+1}} \right\rfloor.
\]
These functions extend naturally to polynomial rings \( R_q \) and \( R_q^k \) by applying them coefficient-wise. For detailed definitions, see \cite{kyber-original}.

\subsubsection{Security and Correctness of Kyber}

Kyber’s encryption scheme ensures Indistinguishability under Chosen Plaintext Attack (IND-CPA) \cite{katz-lindell}, meaning an adversary with access to an encryption oracle cannot distinguish ciphertexts of chosen messages. The public-key encryption (PKE) scheme consists of three algorithms: Kyber.CPA.KeyGen, Kyber.CPA.Enc, and Kyber.CPA.Dec, as defined in \cite{kyber-original}. The parameters include $n=256$, $q=3329$, and module dimensions $k \in \{2,3,4\}$ for different security levels.  

The correctness of Kyber PKE is captured by the  probability $1- \delta$ that Kyber PKE decrypts correctly, where $\delta$ is quantified in \cite{delta-correctness-paper} as:  
\[
        \delta = \Pr\left[ \| \mathbf{e}^T \mathbf{r} + e_2 + c_v - \mathbf{s}^T \mathbf{e}_1 + \mathbf{c}_t^T \mathbf{r} - \mathbf{s}^T \mathbf{c}_u \|_\infty \geq \lfloor q / 4 \rfloor \right].
    \]
The chosen parameters ensure that $\delta < 2^{-128}$.  

To achieve Chosen Ciphertext Attack (CCA) security, Kyber applies the Fujisaki-Okamoto transform to the IND-CPA PKE scheme. The CCA-secure Key Encapsulation Mechanism (KEM) consists of two additional algorithms: Kyber.CCA.Encaps and Kyber.CCA.Decaps \cite{kyber-original}. The encapsulation algorithm generates a ciphertext and a shared secret using hash functions, while decapsulation verifies correctness and ensures security against maliciously constructed ciphertexts. The correctness of Kyber KEM is analyzed in \cite{delta-correctness-paper}, and Kyber KEM is $(1-\delta)$ correct, where:  
$$Pr\left[\text{Decaps}(\text{sk}, c) = K \,\middle|\, (\text{pk}, \text{sk}) \gets \text{KeyGen()}; (K, c) \gets \text{Encaps}(\text{pk})\right] \geq \delta.$$ 
For Kyber512 KEM, $\delta < 2^{-139}$; for Kyber768 KEM, $\delta < 2^{-164}$; and for Kyber1024 KEM, $\delta < 2^{-174}$.

\subsection{Efficient Curvy Protocol}

The Efficient Curvy protocol improves upon the existing pairing-based Curvy protocol by eliminating the use of pairing, replacing it with a much faster, Module-LWE technique. This modification retains the functionality of the original Curvy protocol while significantly improving performance. By utilising the computational efficiency of Module-LWE and retaining the elliptic curve operations, the protocol remains well suited for blockchain applications.

\begin{figure}[H]
  \centering
    \includegraphics[scale=0.63]{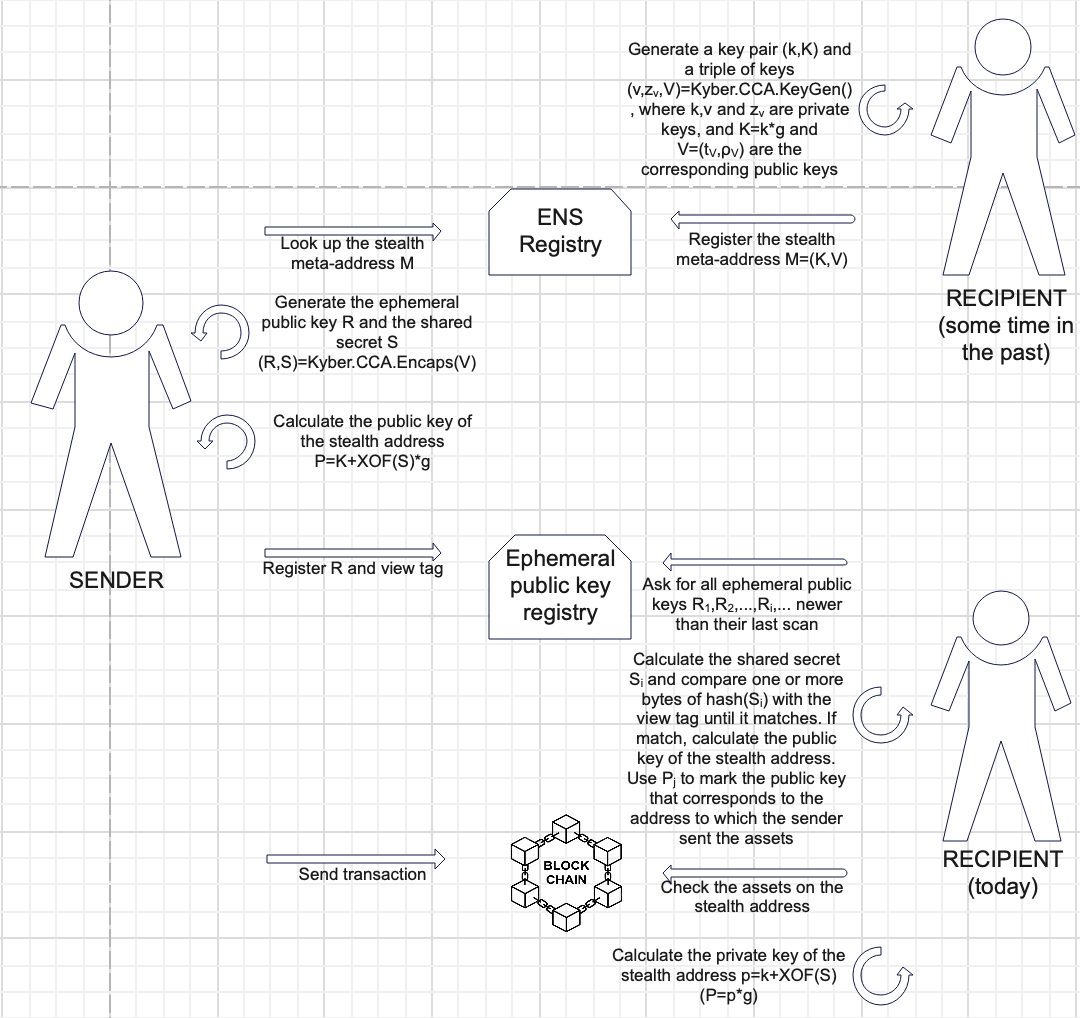}
    \caption{Efficient Curvy Protocol}
    \label{fig:protocol}
\end{figure}

Let $g$ be the generator point of the elliptic curve. Figure \ref{fig:protocol}  illustrates Efficient Curvy protocol, which works as follows: 

$\textbf{1.} $ The recipient generates their spending key $k$ and viewing key $(v, z_v)$ and calculates their stealth meta-address $M=(K, V)$, where $K=k*g$ and $V$ is obtained by calling the function Kyber.CCA.KeyGen.

$\textbf{2.} $ The recipient registers the stealth meta-address by adding an ENS (Ethereum Name Service) record containing the stealth meta-address $M$.

$ \textbf{3.} $ The sender searches for recipient's stealth meta-address $M$ on ENS registry, using recipient's ENS name.

$ \textbf{4.}  $ The sender runs the encapsulation by calling the function Kyber.CCA.Encaps with the public key $V$ as input and derives the shared secret $S$ and the ephemeral public key $R$ from it, i.e.
$$
(R,S)=\text{Kyber.CCA.Encaps}(V).
$$
Note that the sender uses a private randomness generated in the Kyber.CPA.Enc function through the Kyber.CCA.Encaps function.

$ \textbf{5.}  $ The sender calculates the public key of the recipient's stealth address as 
$$P =K + \text{XOF}(S)*g.$$

$ \textbf{6.}  $ The sender publishes their ephemeral public key $R$ to the ephemeral public key registry as well as a view tag (one or more bytes of the hash of $S$).

$ \textbf{7.}  $ The sender sends assets to the recipent's stealth address.

$ \textbf{8.}  $ To discover the stealth address belonging to them, the recipient scans all new ephemeral public keys in the ephemeral public key registry since their last scan. Recipient for each retrieved value $R_i$ calculates the shared secret by calling Kyber.CCA.Decaps function as
$$
S_i = \text{Kyber.CCA.Decaps}(v,z_v,V,R_i).
$$

After that, the recipient compares one or more bytes of the hash of $S_i$ with the view tag until it matches.

$ \textbf{9.}  $ Let's use $R_j$ to mark the ephemeral public keys for which the view tag matches. The recipient calculates the public keys $P_j$ in the same way as the sender. Then the recipient calculates the address and checks if it is the stealth address to which the sender sent the assets.

$ \textbf{10.}  $ The recipient derives the private key that corresponds to the new stealth address. The private key $p$ of the stealth address is
$$p = k + \text{XOF}(S)$$
and can only be calculated by the recipient, as this requires the private key $k$.

The recipient’s stealth address public key \( P \) can only be derived by the sender, who possesses the private randomness \( l \), and by anyone with access to the viewing key \( v \) (such as the recipient or an authorized third party like a tax inspector). Additionally, the private key \( p \) corresponding to the stealth address is directly related to its public key \( P \), since:  
$$
P = K + \text{XOF}(S) * g = (k + \text{XOF}(S)) * g = p * g. $$

\noindent\textbf{Note on scalability:}
Due to the large size of the ephemeral public key $R$, storing it directly on-chain (step 6) can be disproportionately expensive. To solve this problem, the full ephemeral keys $R$ can be stored in an off-chain registry, while only the hashes of these keys are committed on-chain. This significantly reduces gas costs. During scanning (step 8), the recipient can identify relevant entries by comparing the hashes on-chain and then retrieving the corresponding $R_i$ values from the off-chain storage.

\section{Implementation results}

In our implementation of the Efficient Curvy Stealth Address Protocol, we used the Rust programming language along with the KyberKEM library \cite{kyber-kem-repo}. The implementation of our stealth address protocol is available in the GitHub repository \cite{pq-sap-repo}.

Since the most important aspect to optimize in stealth address protocols is the scanning of announcements (ephemeral public keys) by the receiver within the ephemeral public key registry, subsection 5.1 focuses on the comparison of this specific component in our Efficient Curvy protocol with the PQ protocol MLWE SAP from the paper \cite{PQMM}.

In subsection 5.2, we extend the analysis by comparing the scan time of the ephemeral public key registry using our protocol from this paper with the fastest existing non-PQ protocol, in particular ECPDKSAP \cite{MM} - Curvy protocol.

Through these comparisons, we aim to show that our protocol achieves comparable performance to the PQ (MLWE) protocol of \cite{PQMM}, and significantly outperforms the Curvy protocol — while preserving the key property that Efficient SAP remains an Ethereum-friendly protocol.

\subsection{Comparison of MLWE SAP and Efficient Curvy Protocol}

In this experiment, just as in the experiment in subsection 5.2, 10 randomly selected seeds were used to generate private keys and announcements to ensure consistency. The average execution time was recorded and used for comparison.

\begin{figure}[H]
 \centering
   \includegraphics[scale=0.48]{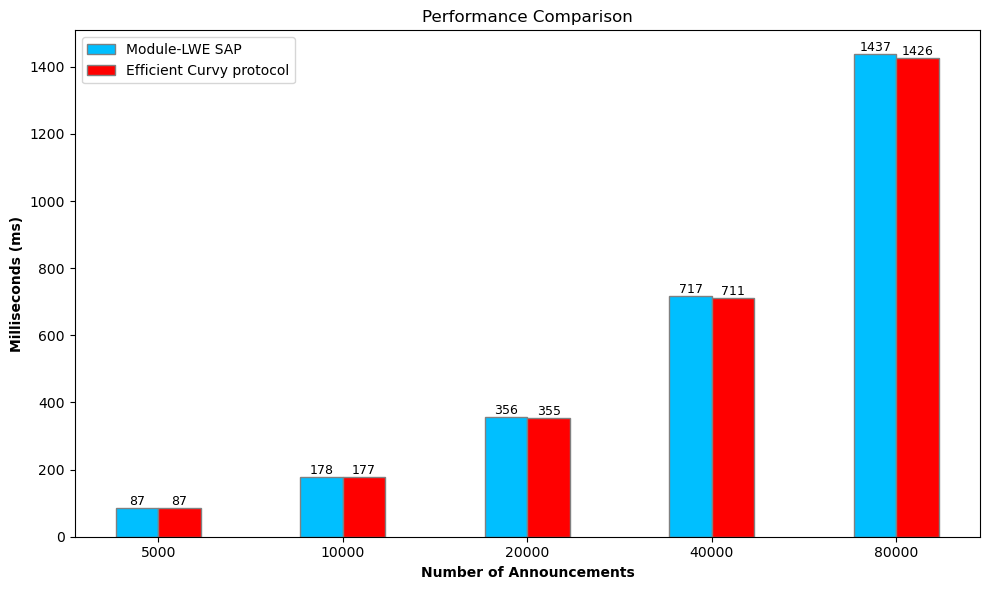}
      \caption{Performance comparison between Module-LWE SAP and Efficient Curvy protocol}
      \label{fig:comparison2}
\end{figure}

Figure \ref{fig:comparison2} presents a comparison of scan times for the ephemeral public key registry with different numbers of announcements (5000, 10000, 20000, 40000 and 80000), comparing the implementation of MLWE SAP from \cite{PQMM} with our implementation of the Efficient Curvy protocol from this paper. For both protocols, the size of the view tag is one byte. Note that the view tag can be larger in both protocols. Increasing the size of the view tag provides more efficiency, but at the cost of writing a larger tag to the registry. If we make a trade-off between the size of the view tag and the efficiency of the mentioned protocols, we come to the conclusion that the one byte view tag is a more than satisfactory choice.

The implementation of Efficient Curvy SAP demonstrates similar efficiency to the referenced implementation of the MLWE SAP protocol (the protocol from \cite{PQMM}, which is the best performing protocol in that paper). For each number of announcments (5000, 10000, 20000, 40000 and 80000), the scanning time of the ephemeral public key registry is almost identical, for both protocols. For comparison, for 80000 announcements, the MLWE SAP protocol is approximately 1\% slower than the Efficient Curvy protocol, with even smaller differences in scanning speed for fewer announcements.

It is important to note that while the MLWE SAP protocol provides post-quantum security, the Efficient Curvy protocol does not, as its security is based on the elliptic curve discrete logarithm problem. However, MLWE SAP is not compatible with Ethereum, meaning its implementation would require the transition from Ethereum to a post-quantum blockchain. In contrast, the Efficient Curvy protocol is inherently Ethereum-friendly and still achieves the same level of efficiency as MLWE SAP.

\subsection{Comparison of Curvy and Efficient Curvy Protocols}

\begin{figure}[H]
  \centering
    \includegraphics[scale=0.48]{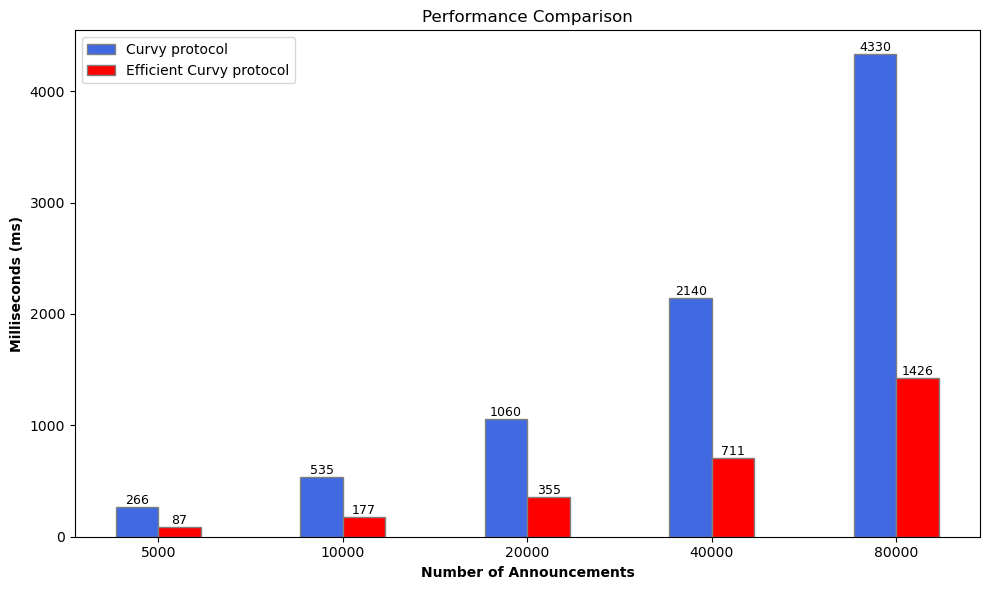}
     \caption{Performance comparison between Curvy and Efficient Curvy protocols}
      \label{fig:comparison1}
\end{figure}

Figure \ref{fig:comparison1} presents a comparison of the scan times for the ephemeral public key registry with different numbers of announcements (5000, 10000, 20000, 40000, and 80000) for both the Curvy and Efficient Curvy protocols. In all cases, the view tag is set to one byte of $\mathrm{hash}(S)$, where $S$ represents the shared secret.

Note that Efficient Curvy protocol gives the best results for all measurements. Specifically, for 80000 announcements, Efficient Curvy protocol outperforms Curvy protocol by 67\%. 

And it is precisely by using the MLWE technique that we have managed to achieve greater efficiency of the Curvy protocol while maintaining the property that the protocol can be used on the Ethereum blockchain. Note that the Efficient Curvy protocol performs registry scanning about 3 times faster than the current Curvy protocol, for all the given numbers of ephemeral keys tested.

\section{Conclusion}

Our main goal in this paper is to provide privacy to the recipient of the transaction using stealth addresses. In this paper, we first gave a description of currently used stealth address techniques, which include scalar multiplication with elliptic curve points, elliptic curve pairing, and PQ techniques such as the Module-LWE technique. Here we introduce a new stealth address protocol that is currently the fastest stealth address protocol on Ethereum. The protocol itself is a hybrid creation between two of our protocols: the Curvy protocol and the MLWE SAP. This protocol does not provide PQ security, but the MLWE technique that we have adapted to the Curvy protocol provides more efficiency in the scanning of ephemeral keys in the registry by the recipient of the transaction. In this way, by combining elliptic curve multiplication and MLWE techniques, we have achieved higher efficiency compared to the Curvy protocol while maintaining the property that the Efficient Curvy protocol can be used on Ethereum.

The efficiency of the Efficient Curvy protocol was evaluated by measuring the time it takes the transaction recipient to compute the public key of its new stealth address. This was compared to both the fastest Ethereum-friendly protocol, the Curvy protocol, and the fastest post-quantum protocol, MLWE SAP. The results clearly show that the Efficient Curvy protocol outperforms the other protocols. It works almost as fast as the MLWE SAP protocol (although Efficient Curvy is slightly faster) and takes about three times less time to scan the ephemeral public key registry compared to the Curvy protocol.

Efficient Curvy is both Ethereum-friendly and open source, making it a strong candidate for real-world applications, especially due to its high efficiency.

\end{document}